\documentclass[letterpaper]{article} 
\usepackage{aaai23}  
\usepackage{times}  
\usepackage{helvet}  
\usepackage{courier}  
\usepackage[hyphens]{url}  
\usepackage{graphicx} 
\usepackage{natbib}  
\usepackage{caption} 
\usepackage{booktabs}
\usepackage{amssymb}
\usepackage{amsmath}
\usepackage{amssymb}
\usepackage{multirow}

\usepackage{xcolor}
\usepackage{colortbl}
\usepackage{placeins}
\usepackage{algorithm}
\usepackage{algorithmic}
\usepackage{newfloat}
\usepackage{listings}
\DeclareCaptionStyle{ruled}{labelfont=normalfont,labelsep=colon,strut=off} 
\floatstyle{ruled}
\newfloat{listing}{tb}{lst}{}
\floatname{listing}{Listing}
\title{Divergences in Following Patterns between Influential Twitter Users and Their Audiences across Dimensions of Identity}
\author {
    Suyash Fulay\textsuperscript{\rm 1},
    Nabeel Gillani\textsuperscript{\rm 1,\rm 2},
    Deb Roy\textsuperscript{\rm 1}
}
\affiliations {
    \textsuperscript{\rm 1}Massachusetts Institute of Technology, Cambridge, MA \\
    \textsuperscript{\rm 2}Northeastern University, Boston, MA\\
    \{sfulay, ngillani, dkroy\}@mit.edu
}

\usepackage{dirtytalk}

\begin{document}

\maketitle

\begin{abstract}

Identity spans multiple dimensions; however, the relative salience of a dimension of identity can vary markedly from person to person.  Furthermore, there is often a difference between one’s internal identity (how salient different aspects of one's identity are to oneself) and external identity (how salient different aspects are to the external world). We attempt to capture the internal and external saliences of different dimensions of identity for influential users (“influencers”) on Twitter using the follow graph. We consider an influencer’s “ego-centric” profile, which is determined by their personal following patterns and is largely in their direct control, and their “audience-centric” profile, which is determined by the following patterns of their audience and is outside of their direct control. Using these following patterns we calculate a corresponding salience metric that quantifies how important a certain dimension of identity is to an individual. We find that relative to their audiences, influencers exhibit more salience in race in their ego-centric profiles and less in religion and politics. One practical application of these findings is to identify \say{bridging} influencers that can connect their sizeable audiences to people from traditionally underheard communities. This could potentially increase the diversity of views audiences are exposed to through a trusted conduit (i.e. an influencer they already follow) and may lead to a greater voice for influencers from communities of color or women.
\end{abstract}

\section{Introduction}

Identity is foundational to who we are and underlies many of our interests and actions \cite{what_is_identity}. It also spans multiple, overlapping dimensions \cite{Jones2000ACM}. The level of importance, or salience, of each of these different dimensions varies from individual to individual and across situations, and is important in order to understand a person's worldview \cite{salience_identity}. Furthermore, identity is both personal and social — and how we see or understand ourselves does not always align with how others see and understand us \cite{int_ext}.
Most of the literature on identity salience is primarily based on in-person surveys with limited samples \cite{identity_as_variable}, and work quantitatively measuring the difference between internal and external identity has been relatively sparse.  Moreover, when identity is studied on social media platforms, these dimensions are often studied in isolation (e.g. there is work on political \cite{political_orientations} and religious \cite{religious_communities} communities separately), but there is less literature that compares the salience of multiple dimensions of identity for the same set of users.

We hope to fill this gap in research by leveraging social media to define and propose new measures of internal and external identity across a range of dimensions.  We believe doing so can not only enhance our ability to understand and compare different notions of identity, at scale, for different types of individuals, but may also have potential to identify users that might act as a “bridge” between audiences and traditionally underheard communities. Additionally, by measuring the salience of all of these dimensions of identity for a single, consistent set of users, we can compare in a common space the relative importance of these dimensions for influencers and their audiences, which to our knowledge has not been done before.

Thus, although internal/external identity can vary significantly between people, we explore whether there are any consistent divergences in these two types of identity, as that may suggest a systematic difference in how the world views these influencers versus how they view themselves. We find that race is most salient for the influencers, while politics and religion are more salient for their audiences. This is an insightful finding for researchers that study identity as it points to a relatively consistent divergence across influencers that may merit further study. Furthermore, we use our proposed metrics to identify influencers that could amplify voices from traditionally underheard communities (e.g. people of color or women) which could be useful for policymakers and influencers alike. For example, we identify influencers with a significant number of followees in communities of color relative to their audiences. Knowing this information, these influencers could use their platform to amplify voices from this community by either retweeting or sharing their content.

\section{Data}
\subsection{Selecting Influencers and Audience Members}
We modify the \say{snowball sampling} approach used in \cite{who_says} to select influencers. We start with a seed set of publicly available celebrity Twitter accounts who are mostly actors, musicians, or politicians \cite{PROFILErehab}. We consider followees two-degrees away from the seed set, and then filter out to those candidates that have over 10K followers and are American. This yielded a set of 12,593 influencers. Next, we select a small random subset ($< 1\%$) of the followers of the influencers and filter to those that follow at least twenty people in our set of influencers. This yielded a set of 80,288 audience members. We test that our results are robust over different samples of audience members.

\subsection{Tagging Dimensions of Identity}

In order to tag each influencer across the different dimensions of identity, we first find the categories associated with a given dimension of identity. For racial/ethnic, LGBTQIA+, and religious categories, we use lists provided by the NIH, the LGBTQIA+ Health Education Center, and Pew respectively \cite{NIH, lgbtq, relig} Then, for each user, we intersect these terms with their Wikipedia categories. We provide the breakdown of categories across dimensions of identity in Table \ref{tab:influencer_categories}.

\begin{table}[]
    \centering
    \bgroup
\def\arraystretch{1}

\begin{tabular}{c p{4.5cm}}    
\toprule
\multicolumn{1}{c}{Identity Dimension} & \multicolumn{1}{c}{Categories}  \\
\midrule
Race& Caucasian (\textit{63}), African-American (\textit{21}), Asian (\textit{7}), Hispanic/Latino (\textit{7}), Native American/Hawaiian (\textit{1})\\
\hline
Gender & Male (\textit{60}), Female (\textit{40}), Non-Binary (\textit{.3})\\
\hline
LGBTQIA+ & None Specified (\textit{91}),
LGBT (\textit{3}),
Gay (\textit{2}),
Bisexual (\textit{1}),
Lesbian (\textit{1}),
Transgender (\textit{0.5}),
Queer (\textit{0.5}),
Pansexual (\textit{0.2}),
Asexual (\textit{0.02})\\
\hline
Religion & None Specified (\textit{78}),
Jewish (\textit{12}),
Christian (\textit{5}),
Catholic (\textit{4}),
Muslim (\textit{1}),
Buddhist (\textit{0.3}),
Hindu (\textit{0.2}),
Atheist (\textit{0.1}),
Sikh (\textit{0.04}),
Spiritual (\textit{0.01})\\
\hline
Political & None Specified (\textit{88}),
Democrat (\textit{7}),
Republican (\textit{4}),
Libertarian (\textit{0.7}),
Independent (\textit{0.5})\\
\bottomrule
\hline
\end{tabular}
\egroup

    \caption{All of the categories used for each dimension of identity, with the percentage of influencers labeled with that category in parenthesis.}
    \label{tab:influencer_categories}
\end{table}

\section{Methodology}
Which influencers users choose to follow on Twitter is related to and indicative of the users' interests \cite{follow_comm}. Therefore, for a given user we construct measures of salience across dimensions of identity based on their followees' identity tags. We acknowledge that this list of identity dimensions is in no way exhaustive, and in fact the precise definition of identity and whether to use socially constructed labels is still being debated \cite{brubaker}. However, these dimensions still provide a useful scaffolding to measure and compare saliences.

After tagging users and creating identity salience metrics, we compute the difference in saliences between influencers and their audiences by utilizing the “social-broadcast” network structure of Twitter. This structure means that a small subset of users entertain large audiences on the platform \cite{who_says}.  We can gauge how an influencer's audience views them in the broader cultural landscape by investigating who else the audience follows, while measuring the dimensions the influencer deems most salient by understanding their own followees. We formalize this concept in the “audience” and “ego” centric social media profiles for influencers.  Since internal identity is how one views oneself, and following choices are shown to reflect individual choices and preferences \cite{follow_comm}, we view the \say{ego} centric profile as a measurable analog to internal identity. Conversely, since audience following patterns are more a reflection of how the rest of the world views an influencer, we consider the \say{audience} centric profile to be a similar concept to external identity.

\subsection{General Notation}
Let $M=  12593$ be the number of influencers in our set, with individuals $K = \{k_1...k_M\}$, and $N = 80288$ be the number of audience members comprised of individuals $A = \{a_{M+1}...a_{M+N}\}$. These are two distinct sets, so $A \cap K = \emptyset$ We define an matrix $F \in \mathbb{R}^{ M \times (M+N)}$ where 
\begin{equation}
  F_{i,j} =
  \begin{cases}
    1 & j \leq M,  k_j\text{ follows } k_i \\
    1 & j > M,  a_j\text{ follows } k_i \\
    0 & \text{o.w.}
  \end{cases}
\end{equation}
This is effectively the \say{followee matrix}. Each column $j$ corresponds to an individual (either an audience member or an influential user), and each entry in the column is one if user $j$ follows the influential user represented in the $i$th row, and zero otherwise. The vector of the followees of individual $j$ will be denoted by the $j$th column of $F$ as $\mathbf{f_j} \in \mathbb{R}^{M \times 1}$.

\begin{figure*}
    \centering
    \includegraphics[scale=.82]{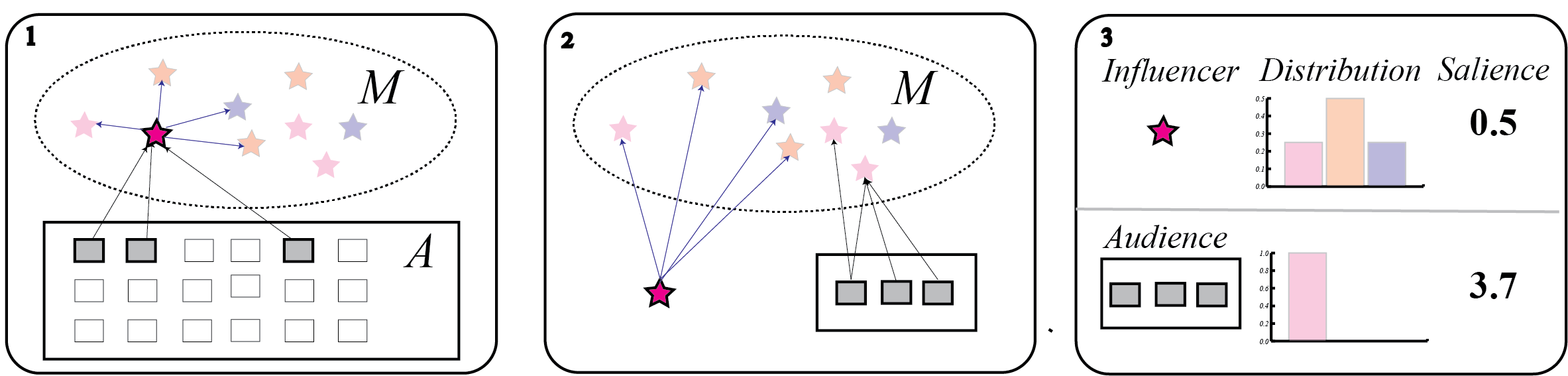}
    \caption{ An example of computing identity salience. (1) shows the universe of influencers $M$ with the set of audience members $A$. We focus on a single influencer (the dark outlined pink star) and their followers (the dark grey boxes in $A$), with arrows indicating a follow relationship. (2) illustrates how we find the influencer's followees in $M$ as well as their audience's followees. Finally, given those follow relationships we find a distribution over identity dimensions and compute a salience metric in (3). Here, since the audience's followees are more homophilous than the influencer's, the audience would have a higher salience score on this dimension of identity.}
    \label{fig:my_label}
\end{figure*}

\subsection{Race and Gender}
We use an entropy based metric to estimate the salience of these dimensions of identity to the individual. Salience of racial identity has been associated with homophily in relationships for Caucasians, Hispanics, and African-Americans \cite{race_homophily}, and high gender salience in women is linked to increased likelihood of friendship with another woman interlocutor \cite{gender_sal}. Thus, we assume that if a user's followees tend to be of a particular race or gender, this is likely an indication that this is a relatively important part of the user's identity.

\subsubsection{Notation}

We define binary matrices $D^{race}$ and $D^{gender}$ in $\mathbb{R}^{6 \times M}$ and $\mathbb{R}^{3 \times M}$ respectively. In $D^{race}$ each row corresponds to a different racial category, while in $D^{gender}$ each row corresponds to a different gender category. 

\begin{equation}
  D^{race}_{c,i} =
  \begin{cases}
    1 & k_i\text{ is tagged as having racial category } c \\
    0 & \text{o.w.}
  \end{cases}
\end{equation}
$D^{gender}$ is defined analogously.
\subsubsection{Metric}
We  calculate the ego-centric identity salience score $e^{race}_{j}$ for an individual with friends defined by $\mathbf{f_j}$ as follows 
\begin{equation}
    \label{norm_vec}
    \mathbf{v^{\text{race}}_{j}} = \frac{{D^{\text{race}}}\mathbf{f_j}}{||D^{\text{race}}\mathbf{f_j}||_1}
\end{equation}
Each entry in this vector is the percentage of user $j$'s followees that have the corresponding racial category $c$. We let $v^{race}_{j, c}$ be the $c$th entry of \eqref{norm_vec}. We calculate the entropy of this normalized vector  to get our metric 
\begin{equation}
    w^{\text{race}}_{j} =- \sum_{c=1}^{6} v_{j,c}^{\text{race}} \log(v_{j,c}^{\text{race}})
\end{equation}
This value is minimized when all the followees of an individual have the same tag, and maximized when the distribution over followees' tags is uniform.  We then normalize this metric across all users to have mean zero and variance one, and then multiply by negative one so that lower entropy corresponds to higher salience.

\begin{equation}
    e_{j}^{race} = -\frac{ w_{j}^{\text{race}} -  \mu(w_1^{\text{race}}...w_{M+N}^{\text{race}})}{\sigma(w_1^{\text{race}}...w_{M+N}^{\text{race}})}
\end{equation}
where $\mu$ and $\sigma$ compute the mean and standard deviation across all users for the given dimension of identity. The metric is defined analogously for gender.

\subsubsection{Ego and Audience Centric Identity Salience Scores}
For an influential user $k_j$, their ego-centric identity score is simply $e_{j}^{\text{dim}}$ where $\text{dim} \in \{\text{race}, \text{gender}\}$. Their audience-centric identity score $a_{j}^{\text{dim}}$ is the average of the $e_{j}^{\text{dim}}$s for each of their followers in $A$.

\subsection{LGBTQIA+ Status, Political Orientation, and Religious Affiliation}
Since tags on LGBTQIA+ status, political orientation, and religious affiliation are not inferable from every Wikipedia page, we take a different approach to computing salience. The presence of these tags often indicate a publicly known aspect of a person's identity. Therefore, we gauge user salience along these dimensions of identity by the number of followees that are tagged as having a category in that dimension. This approach is similar to Lim and Datta, who tag influencers with media categories to determine user interest in a similar fashion.

\subsubsection{Metric}

To capture the salience of one of these dimensions of identity for an individual, we calculate the percentage of their followees without a tag in that dimension. Let $f_j^{\text{-religion}}$ be the number of followees of a user \textit{without} a religious tag in $K$ and $f_j^{\text{total}}$ be the total number of followees they have in $K$. Then our metric is simply

\begin{equation}
    w_{j}^{\text{religion}} = \frac{f_j^{\text{-religion}}}{f_j^{\text{total}}}
\end{equation}
Similar to the metric for race and gender, we normalize across all users and multiply by negative one to get $e_j^{\text{religion}}$. This metric is maximized when all of a user's followees have a tag for a given dimension of identity and is minimized when none of them have a tag. The audience centric metric is calculated analogously to race/gender.

\section{Results}

\begin{figure}
    \centering
    \includegraphics[scale=.485]{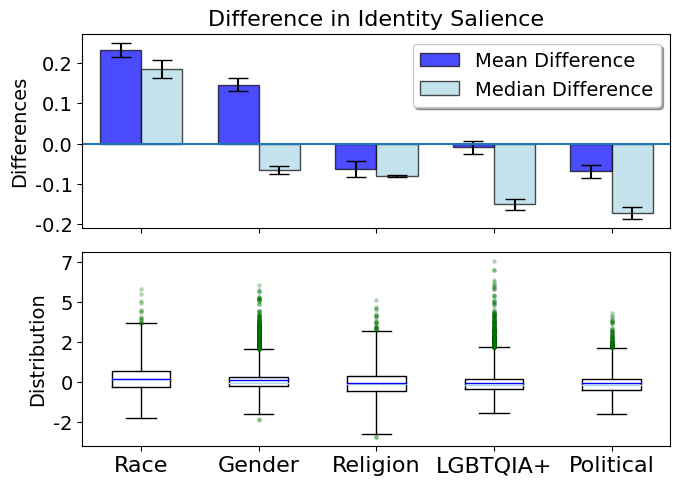}
    \caption{The top chart shows the average difference in ego and audience centric profiles ($\frac{1}{M}\sum_{i=1}^{M} e_i^{\text{dim}} - a_i^{\text{dim}}$) and the median difference. The intervals shown are from a bootstrapped confidence intervals with a $99\%$ confidence level. The box plot  illustrates the general right skew of the data, especially along the gender and LGBTQIA+ dimensions.}
    \label{fig:dist}
\end{figure}
\subsection{Estimating the Magnitude and Significance of Divergences in Profiles}
After computing the above metrics, we are primarily interested in investigating the difference in the ego and audience centric profiles. We construct the set of differences $D^{\text{dim}} = \{e^{\text{dim}}_j - a^{\text{dim}}_j \mid j \in [0,M] \}$ for each dimension of identity and run several tests to determine any systematic divergences.

We find that racial salience is higher for influencers, while religion and politics is more salient for their audiences. As illustrated in Figure \ref{fig:dist}, there are are a set of influencers that seem to have extremely high salience for gender/LGBTQIA+ relative to their audiences, which causes the mean and median differences to diverge significantly. To test the significance of the results, we ran a paired t-test \cite{ttest}, the Wilcoxen signed rank test \cite{wilcoxen}, and a bootstrap approach with paired samples. The latter two tests are non-parametric, meaning they do not depend on the Gaussianity of the distribution of differences. We use the Bonferonni correction \cite{bonf} to correct for the chance of a Type I error (p-values multiplied by 15) and find that race, religion, and political differences are still significant ($p << .01$).

While it is difficult to know precisely why we see these divergences, one potential reason that we see more homophily along race for influencers relative to audiences is that influencers often utilize Twitter more professionally, and occupation tends to be related to race \cite{race_occ}. When investigating the influencers with extremely high gender salience, we find that a significant number of them are football players that follow other football players, who are almost exclusively men. Thus, a consequence of occupational homophily for Twitter influencers could lead to the high racial/gender salience we observe. Additionally, since an influencer's followees are public, they may be more careful when following political elites as to not alienate any of their audience. However, we emphasize that a current limitation is that we cannot explain concretely the reason for these differences and would encourage future work in this direction.

\begin{table}[!htbp]
    \centering
    \setlength{\aboverulesep}{0pt}
    \setlength{\belowrulesep}{0pt}
    
\begin{tabular}{lccccc}
\toprule
\multicolumn{1}{c}{\multirow{2}{*}{Inf.}}  & \multirow{2}{1cm}{Salience Diff} & \multirow{2}{1cm}{Bridge to} & \multicolumn{2}{c}{\% Followees}\\
\cmidrule(lr){4-5}
   & & & Inf. & Aud.
\\
\midrule
Dolly Parton  
& 3.32 &\multirow{3}{*}{Women}&  94\% & 43\% \\
      Ingrid Nilsen  & 1.00 & & 76\% & 49\% \\
   Melinda Gates  & .68 & & 73\% & 41\% \\

\hline
Allen Iverson 
& 2.40 &\multirow{3}{*}{POC} & 77\% & 60\% \\
  Kendrick Lamar  & 1.19& &86\% &57\% \\
   Katrina Taylor  & .97& &84\% &66\% \\ 

\bottomrule
\end{tabular}
    \caption{We identify influencers that can act as \say{bridges} from historically underheard communities to their audiences using our metric, and contextualize the gap in identity salience between the influencers and their audiences. For example, along the dimension of gender Dolly Parton exhibits significantly more salience (2.9) than her audience (-.4) leading to a difference of 3.3. Investigating this gap further, we see that 94\% of Dolly Parton's followees are women, whereas on average her audience's followees are only 43\% women.}
    \label{tab:bridges}
\end{table}

\subsection{Identifying \say{Bridging} Influencers}
One additional application of these metrics is to identify \say{bridging} influencers that connect their audiences to voices from communities to which they are underexposed. To find these users, we examine influencers that exhibit a large divergence from their audience in salience across gender and race. Then, we determine how their following patterns differ from their audiences and highlight cases where they could act to amplify messages from traditionally underheard communities. In Table \ref{tab:bridges}, we list a few example influencers and focus particularly on those that can act as bridges to women or people of color (POC). If these influencers were aware of the unique role they could play for their audiences, they may be more intentional about retweeting or sharing content from the underheard communities that they follow, but their audience does not.
Thus, in addition to the sociological findings on divergences in internal/external identity salience, we also see the potential to leverage these insights into collaborations with influencers to connect their audiences to underheard communities.

\section{Discussion and Limitations}
Although there are many factors that influence following decisions, these results suggests a systematic divergence in \say{internal} versus \say{external} salience on these dimensions of identity. We quantify the extent of this discrepancy, with race and politics being the most divergent. Moreover, since who one follows on Twitter partially determines the type of information one receives on the platform, it is also an important result that the influencers' followees (and thus whose content they see) tend to be more homophilous along race than their own audiences. This information could even be given to users interested in their own following patterns to encourage increasing their diversity of followees \cite{gillani2018me} or expanding awareness to other communities. Influencers could also benefit from seeing their audience's most salient dimensions of identity to better connect with them and understand how they fit in to their audience's broader following patterns. We also acknowledge the limitations of the current work. Since this analysis only includes five dimensions of identity, it could be affected by other latent factors and we make no causal claims on reasons for the observed divergences. Additionally, any systemic bias of which users have Wikipedia pages or which categories are present or omitted could influence the results. However, we believe that this work should enrich our notion of the nuance of identity and provide new methods for exploring and understanding identity on Twitter.

\bibliography{aaai23}

\end{document}